\documentclass[prl,aps,twocolumn,showpacs]{revtex4}

\usepackage{epsfig}

\begin{document}

\title{ $^{77}$Se NMR study of pairing symmetry and spin dynamics in 
K$_{y}$Fe$_{2-x}$Se$_2$}

\author{Weiqiang Yu}
\email{wqyu_phy@ruc.edu.cn}
\author{L. Ma}
\author{J. B. He}
\author{D. M. Wang}
\author{T.-L. Xia}
\author{G. F. Chen}
\author{Wei Bao}

\affiliation{Department of Physics, Renmin University of China, Beijing 
100872, China}

\date{\today}


\begin{abstract}

We present a $^{77}$Se NMR study of the newly discovered iron selenide superconductor 
K$_{y}$Fe$_{2-x}$Se$_2$, in which $T_c = 32$ K. Below $T_c$, the Knight 
shift $^{77}K$ drops sharply with temperature, providing strong evidence 
for singlet pairing. Above $T_c$, Korringa-type relaxation indicates 
Fermi-liquid behavior. Our experimental results set 
strict constraints on the nature of possible theories for the mechanism of 
high-$T_c$ superconductivity in this iron selenide system. 
  
\end{abstract}

\maketitle

The discovery of iron pnictide superconductors \cite{Hosono_Jacs_130_3296, 
Chen_Nature_453_761, Chen_PRL_100_247002, Ren_MRI_12_105} has attracted 
intense research interest. The two most prominent features of high-temperature 
superconductivity in these systems are the following. First, the iron 
pnictide superconductors originate from antiferromagnetic semimetals 
\cite{Chen_PRL_100_247002}, with multiple electron and hole Fermi surfaces 
which are gapped in the superconducting state  
\cite{Ding_EPL_83_47001, ZhouXJ_CPL}. Second, there are pronounced spin 
fluctuations \cite{Imai_prl_102_177005, Kitagawa_PRL, Nakai_PRL_2010}, 
which are gapped and condense into a spin resonance mode 
\cite{Dai_Nature_453_899, Christianson_Nature, Lumsden, Bao_FeSe}. These 
observations suggest that both magnetic correlations and interband transitions between the Fermi 
surfaces are essential intrinsic properties, 
and support an $s^\pm$ gap symmetry in the superconducting state 
\cite{Mazin_PRL_101_057003, Kuroki_PRL_101_087004}. 

High-temperature superconductivity has also been observed very recently in 
the ternary iron selenides $A$$_{y}$Fe$_{2-x}$Se$_2$ ($A$ = K, Rb, Cs, \dots) 
\cite{Guo_PRB_82_180520, Mizuguchi_10124950, Chen_CM_10125525, 
FangMH_CM_10125236, Chen_CM_11010789}. The transition temperature, $T_c\sim$ 
32 K, is very much higher than that in the binary system FeSe at the ambient 
pressure \cite{WuMK}, and is surprising because the system is thought to be 
heavily electron-doped. Initial angle-resolved photoemission (ARPES) studies 
\cite{Ding_CM_10126017, Feng_CM_10125980} revealed only one or more large 
electron bands on the Fermi surface, while a hole band is yet to be confirmed. Theoretically, 
a band insulator \cite{Shein_CM_10125164}, possibly with magnetic correlation
effects \cite{Xiang_CM_10125536}, has been suggested for the stoichiometric 
compound $A$Fe$_2$Se$_2$. However, iron-vacancy order affects both the 
normal-state metallic behavior and the superconductivity in 
$A$$_{y}$Fe$_{2-x}$Se$_2$ very strongly \cite{FangMH_CM_10125236, 
Chen_CM_11010789}. These facts immediately raise the question of whether 
the superconducting gap symmetry and magnetic correlations in this newest 
family of iron selenide superconductors are in fact quite different from 
the pnictides and the Fe(Se,Te) systems.  

Nuclear Magnetic Resonance (NMR) is a bulk probe, which is extremely 
sensitive to the low-energy excitations associated with electronic 
correlation effects, and thus to the pairing symmetry of superconducting
states. The Knight shift, $K$, probes the static electron susceptibility 
$\chi(q)$ at $q = 0$, while the spin-lattice relaxation rate $1/T_1$ measures 
a weighted sum of $\chi''(q)$ over all momenta. We have performed $^{77}$Se 
Knight-shift and $1/^{77}T_1$ measurements on single crystals of 
K$_{y}$Fe$_{2-x}$Se$_2$. The Knight shift we observe provides direct 
evidence for singlet pairing symmetry in K$_{y}$Fe$_{2-x}$Se$_2$. However, 
the Hebel-Slichter coherence peak in $1/^{77}T_1$ is not present at the 
field (11.764 T) of our measurements. We find Korringa-type behavior in 
$^{77}K$ and $1/^{77}T_1$ above $T_c$, suggesting that the superconducting 
state forms from a Fermi-liquid state.

\begin{figure}
\includegraphics[width=9cm, height=8cm]{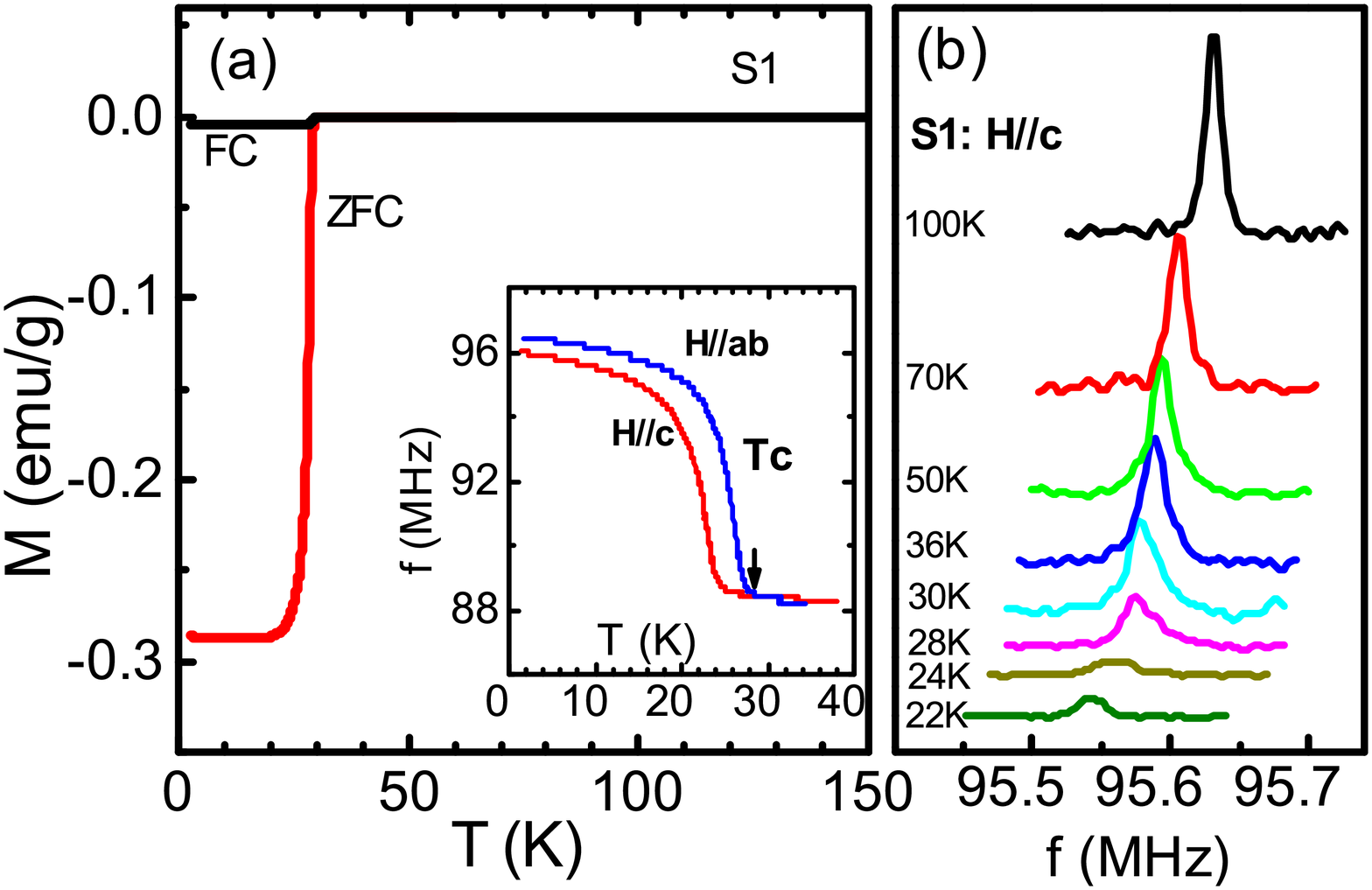}
\caption{\label{sus1}(color online) (a) DC magnetization of 
K$_{0.82}$Fe$_{1.63}$Se$_2$ single crystal (sample S1), measured in a 10 
Oe field under field-cooled (FC) and zero-field-cooled (ZFC) conditions. 
Inset: tuning frequency $f$ of the NMR coil in a magnetic field of 11.764 
T. The frequency shift below $T_c$ depends on the AC susceptibility, 
$\Delta f(T) \propto - \chi_{ac}(T)$. Differing results for fields 
oriented along the $c$-axis or in the $ab$-plane indicate an anisotropic 
$H_{c2}$. (b) $^{77}$Se NMR spectra at different temperatures. The spectra 
broaden below $T_c$ and shift to lower frequencies.}
\end{figure}

Single crystals K$_{y}$Fe$_{2-x}$Se$_2$, with $T_c \approx 30$ K, 
were synthesized by the Bridgeman method \cite{Chen_CM_11010789}. 
For our NMR study, two single crystals with approximate dimensions 
$\sim$3$\times$2$\times$0.5mm$^3$ were chosen. The determination of exact 
stoichiometries in the A$_{y}$Fe$_{2-x}$Se$_2$ series has been found, after 
several measurements by different techniques to be a complex and subtle 
issue. For an accurate characterization of our samples, we have used 
inductively coupled plasma (ICP) measurements, which determined sample S1 
to have composition K$_{0.82}$Fe$_{1.63}$Se$_2$ and to show a monotonic 
increase of resistivity with temperature up to 300 K. Sample S2 was 
determined to have stoichiometry K$_{0.86}$Fe$_{1.62}$Se$_2$ and to show a 
resistivity peak around 150 K \cite{Chen_CM_11010789}. Bulk superconductivity 
was confirmed by measurements of DC magnetic susceptibility [Fig.~\ref{sus1}(a)] 
and heat capacity \cite{Zeng_CM_11015117}. Transmission Electron Microscopy 
(TEM) and X-ray diffraction measurements  \cite{LiJQ_CM_11012059, 
Zavdlij_CM_11014882} have suggested a bulk ordering of Fe vacancies 
in crystals grown under the same conditions. Enlargement of the crystalline 
unit cell, consistent with such ordering, is supported by the observation 
of many additional phonon modes observed in Raman-scattering 
\cite{Zhang_CM_11012168} and infrared studies \cite{WangNL_CM_11010572}. 

Sample S1 was placed on a rotator, which allows the orientation of the NMR 
field (11.764 T) to be changed from the crystalline $ab$-plane to the 
$c$-axis. The superconducting transition was monitored {\it in situ} by the 
high-frequency response of the NMR coil, and $T_c$ determined from the sharp 
increase in frequency upon cooling [inset Fig.~\ref{sus1}(a)]. For S1, $T_c$ 
is higher for fields in the $ab$-plane than for field along the $c$-axis. 
This anisotropic suppression of $T_c$ in the NMR field is consistent with 
the results of Ref.~\cite{Chen_CM_11010789}. S2 was ground into a coarse 
powder to gain better RF penetration below $T_c$. The Knight shift 
$^{77}K(T)$ is obtained from $K(T) = (f - \gamma_n B)/\gamma _n B$, where 
$^{77}\gamma_n = 8.118$ T/MHz is the gyromagnetic ratio of $^{77}$Se and 
$f$ is the measured resonance frequency in the magnetic field $B$. The 
relaxation rate $1/^{77}T_1$ was obtained by the inversion-recovery method. 
The $^{77}$Se spectra of S1, with the field oriented along the $c$-axis, 
are shown for different temperatures in Fig.~\ref{sus1}(b). The line width  
$\Gamma \sim$18 kHz observed above $T_c$ is narrow, changes little with 
temperature, and has the same value within our resolution for both field 
orientations. The spectra of S2 have $\Gamma \sim$14 kHz (data not shown). 
This narrow line width and the excellent fit to the spin recovery by a 
single exponential function for the $^{77}T_1$ measurements indicate that 
our NMR measurements were performed on a single phase. There is no 
evidence for qualitative differences in $^{77}K$ and $1/^{77}T_1$ between the 
two samples.  

\begin{figure}
\includegraphics[width=9cm, height=8cm]{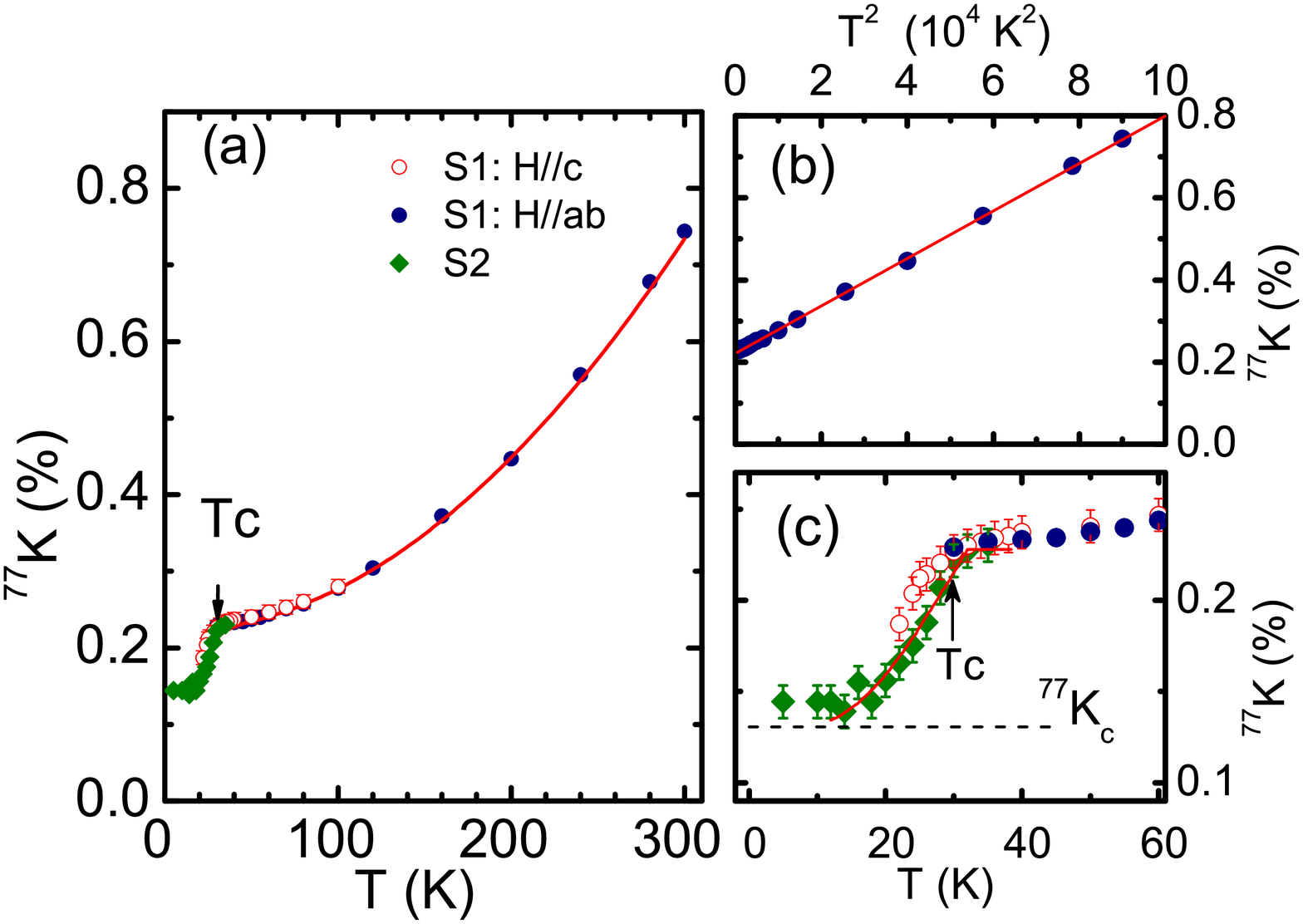}
\caption{\label{knight2}(color online) (a) Temperature dependence of 
the $^{77}$Se Knight shift for the two samples. The solid line is a 
guide to the eye. (b) $^{77}K$ shown as a function of $T^2$; the solid 
line is a fit $K(T) = a + b T^2$. (c) Low-temperature behavior of $^{77}K$; 
the solid line is a guide to the eye. The dotted line denotes approximately 
the chemical contribution to the Knight shift ($^{77}K_c$, see text).} 
\end{figure}

Figure \ref{knight2}(a) displays our Knight-shift data. For S1, measurements 
were performed from 30 K to 300 K with the field in the $ab$-plane and from 
22 K to 100 K with field along the $c$-axis. For S2, measurements were 
performed from 2 K to 35 K. Above $T_c$, $^{77}K$ increases with temperature, 
and in the overlapping temperature range we find no appreciable difference 
between the two field orientations. This indicates an isotropic Knight shift 
above $T_c$. Figure \ref{knight2}(c) shows the detailed features of $^{77}K(T)$
at low temperatures, where it falls abruptly from $T_c$ down to $T_c/2$. 
Below $T_c/2$, however, $^{77}K(T)$ changes little. We suggest that the behavior
below $T_c/2$ is governed by vortex-core contributions, because the RF 
screening effect is strong in the superconducting regions. Here we neglect
the superconducting diamagnetic shift, which has been found \cite{Ma_11023888}
to be very small in another superconductor from the same structural family
of iron selenides. We will discuss the determination of the chemical
shift $K_c$ in Fig.~\ref{knight2}(c) below.  Concerning temperatures directly below $T_c$, the sharp 
drop in $^{77}K(T)$ provides direct evidence for singlet superconductivity 
in K$_{y}$Fe$_{2-x}$Se$_2$. This result places a strong constraint on the 
pairing mechanism, as the gap symmetry must be $s$-wave, $d$-wave, or some 
other even orbital symmetry of the Cooper pairs. Taken together with 
definitive ARPES measurements on high-quality surfaces, our results will 
have fundamental implications for understanding the newest family of
iron-based superconductors.  

\begin{figure}
\includegraphics[width=9cm, height=7cm]{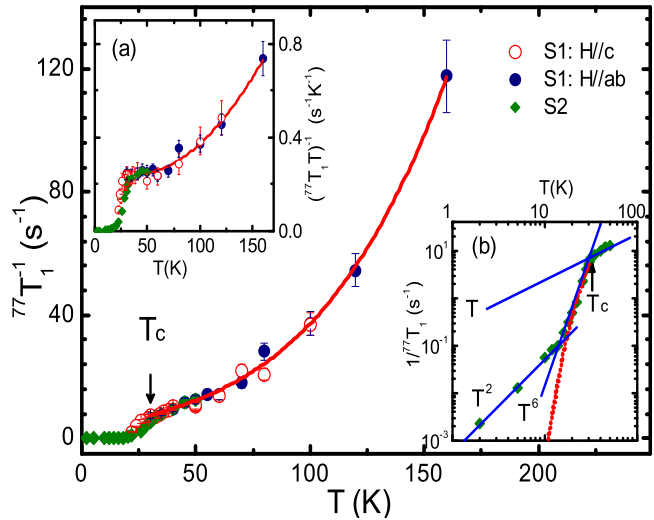}
\caption{\label{slrr3}(color online) Temperature-dependence of the 
spin-lattice rate $1/^{77}T_1$ for both samples. Inset (a): data shown as 
$1/^{77}T_1 T$. Inset(b): $1/^{77}T_1$ shown on logarithmic axes at low 
temperatures. Solid lines indicate different types of power-law dependence
($1/T_1 \sim T^n$), while the dotted line is a fit to an activated form 
$1/^{77}T_1 = A \exp (-\Delta /k_B T)$ (see text).}
\end{figure}

To gain further insight into the gap symmetry, we have measured the 
spin-lattice relaxation rate of $^{77}$Se, shown in Fig.~\ref{slrr3}. 
For S1, data was taken from 30 K to 140 K with the field in the $ab$-plane 
and from 22 K to 100 K with field along the $c$-axis, while for S2 we 
measured from 2 K to 50 K. $1/^{77}T_{1}$ decreases with temperature, 
showing a sharp drop below $T_c$. From Fig.~\ref{slrr3}(b), which focuses 
on the low-temperature regime, it is clear that there is no Hebel-Slichter 
coherence peak of the type usually associated with conventional $s$-wave 
superconductivity \cite{Hebel_PR_113_1504-1519, Hebel_PR_116_79}. For the 
convenience of the reader, we represent by straight lines the functional 
forms $1/^{77}T_1 \sim T$, $1/^{77}T_1 \sim T^2$, and $1/^{77}T_1 \sim T^6$. 
If $1/^{77}T_1$ were determined by a single, constant energy gap, one would 
expect $1/T_1 \sim e^{-\Delta/k_B T}$. Taking the value $\Delta \approx 10.3$ 
meV from the ARPES measurements of Ref.~\cite{Feng_CM_10125980}, the 
activation curve offers a reasonable fit to our data only in the region 
from $T_c$ down to approximately $T_c/2$.

The absence of a coherence peak in cuprate superconductors is interpreted 
as evidence for $d$-wave pairing. Such observations have been reported in
many iron-based superconductors \cite{Grafe_PRL_101_047003, 
Matano_EPL_83_57001, Sato_JPSJ, Kobayashi_JPSJ_78_073704, Zhang_prb_81,
Fukazawa_JPSJ_78_033704, Hammerath_CM_09123681}, and has been interpreted 
as a indication of $s^\pm$ symmetry on separate Fermi surfaces 
\cite{Parker_PRB_78_134524, Chubukov_PRB_78_134512, Parish_PRB_78_144514, 
Bang_PRB_79_054529, Nagai_08091197}. However, the orbital symmetries of 
these studies may not be applicable in K$_{y}$Fe$_{2-x}$Se$_2$, where the 
presence of a hole band at the center of the Brillouin zone has not been 
confirmed despite a number of ARPES investigations. Further, a nodeless 
gap has been suggested in one ARPES study \cite{Feng_CM_10125980}. We note 
here that our data are obtained in a high magnetic field, and thus that 
measurements at much lower fields will be required to exclude the possibility 
that K$_{y}$Fe$_{2-x}$Se$_2$ may in fact be a system with a field-suppressed 
coherence peak, as has been found in $A_3$C$_{60}$ superconductors 
\cite{Stenger}.

\begin{figure}
\includegraphics[width=8cm, height=7cm]{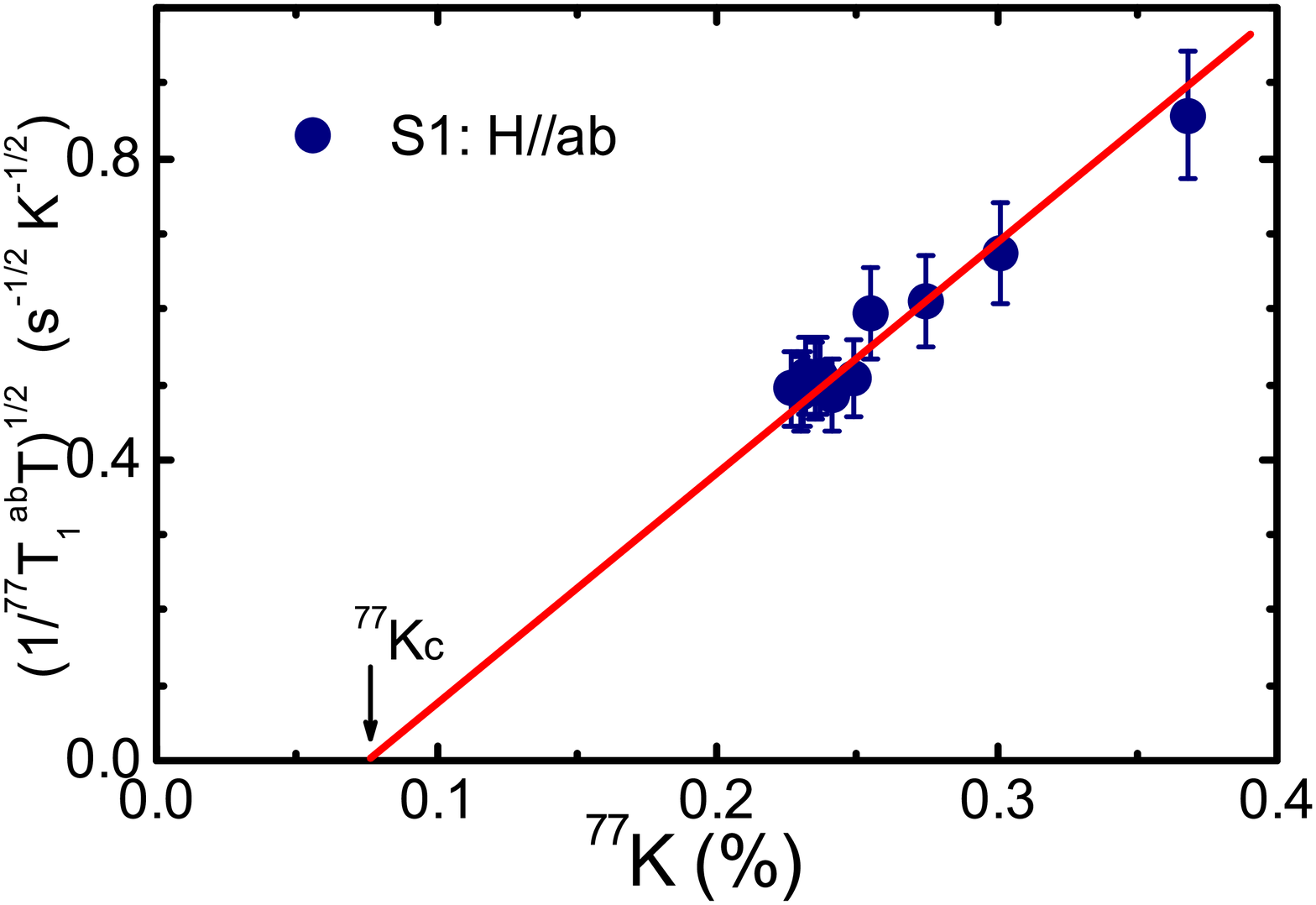}
\caption{\label{korringa4}(color online) Korringa plot of 
$(1/^{77}T_1T)^{0.5}$ against $^{77}K(T)$ with temperature as an 
implicit parameter. The solid line represents the Korringa relation.}
\end{figure}

We turn next to $1/^{77}T_1$ in the normal state. As shown in 
Fig.~\ref{slrr3}, $1/^{77}T_1$ is isotropic with field orientation and 
increases with temperature above $T_c$. In inset (a) we show $1/^{77}T_1 
T$ for both samples. This quantity is almost $T$-independent over a 
substantial range above $T_c$, before increasing rapidly at higher 
temperatures. For a Fermi liquid, $1/T_1 T$ is a constant depending 
on the density of states at the Fermi level. Further, a Fermi liquid 
satisfies the Korringa relation, $1/(T_1T)^{0.5}\propto K_s$, where the 
spin part of the Knight shift is given by $K_s(T) = K(T) - K_c$ and the 
chemical shift $K_c$ is temperature-independent. We show in 
Fig.~\ref{korringa4} the relation between $1/(^{77}T_1T)^{0.5}$ and 
$^{77}K(T)$ for S1 in the temperature range between 30 K and 140 K. 
The linear dependence demanded by the Korringa relation is obvious. 
The gradient $k$ in Fig.~\ref{korringa4} appears in the Korringa 
ratio, $S = (4\pi k_B/\hbar)(\gamma _n/\gamma _e)^2/k^2 = (4\pi k_B/\hbar) 
(\gamma _n/\gamma _e)^2 T_1 T K^2_s$. From our data, $S\approx 0.8-1.5$ 
between 30 K and 140 K, which falls in the range for a canonical Fermi 
liquid with no strong ferromagnetic ($S\gg 1$)  or antiferromagnetic 
($S\ll 1$) spin fluctuations \cite{Nakai_PRL_101_077006, Jelic_LiFeAs_NMR, 
Li_JPSJ_79}. Our data therefore indicate strongly a rather conventional 
Fermi-liquid state for the conduction electrons in K$_{y}$Fe$_{2-x}$Se$_2$
above $T_c$. 

Before leaving Fig.~\ref{korringa4}, we comment briefly that the $K$-axis 
intercept yields $^{77}K_c \approx 0.08\%$. Usually $K_c$ can be obtained 
directly in the superconducting state, where $K_s(T)$ approaches zero as 
T $\rightarrow$ 0. In Fig.~\ref{knight2}(c), the flattening of $^{77}K(T)$ 
below 15 K indicates a higher value, $^{77}K_c \approx 0.13\%$; this latter 
method may, however, overestimate $^{77}K_c$ because $^{77}K_s(0)$ can be 
enhanced by the high NMR field used in our experiments.

Above $T_c$, activated increases of $K$ and $1/T_1T$ with increasing 
temperature have been observed in some iron pnictide superconductors, 
and have been interpreted as a pseudogap state \cite{Ning_PRL_104}. 
We stress that our $^{77}K$ data for K$_{y}$Fe$_{2-x}$Se$_2$
can be fitted by the simple function
$K(T) = a + b T^2$ up to 300 K [Fig.~\ref{knight2}(b)], with 
parameters $a \approx 0.23\%$ and $b \approx 6\times 10^{-6}\%/$K$^2$. 
It is currently not clear whether this increase of $^{77}K$ with 
temperature can be caused by a band effect or by a magnetic correlation 
effect \cite{Ning_PRL_104, Zhang_EPL}. The Knight shift of $0.74\%$ at 
room temperature is to our knowledge the largest yet measured among the 
iron-based superconductors 

In the binary iron superconductor FeSe, strong antiferromagnetic 
spin fluctuations, with a Curie-Weiss upturn in $1/T_1T\sim 1/(T+\Theta)$, 
are indicated \cite{Imai_prl_102_177005}. It has been argued that these 
spin fluctuations are responsible for superconductivity. A similar 
Curie-Weiss upturn in $1/T_1T$ has also been reported in many iron 
pnictide superconductors, \cite{Kita_JPSJ_77_114709, Baek_PRL,
Yashima_JPSJ_NMR, Ning_PRL_104, Nakai_PRL_2010, Urbano_PRL, Ma_prb_2010}. 
These are in contrast to our K$_{y}$Fe$_{2-x}$Se$_2$ samples, where 
no upturn of $1/^{77}T_1T$ is detected. However, we caution that our 
measurements cannot exclude antiferromagnetic spin fluctuations at the 
wave vector $q = (\pi,\pi)$, because these would cancel at the Se 
site due to its location at the center of the Fe square 
\cite{Nakai_PRL_101_077006}. We note in this context that low-energy 
spin fluctuations are also rather weak in the overdoped LaFeAsO$_{1-x}$F$_x$ 
compound, while $T_c$ is reasonably high ($T_c \approx$ 40 K at 3 GPa pressure) 
\cite{Nakano_PRB_82_172502}.

In summary, we have performed $^{77}$Se Knight shift and $1/^{77}T_1$ 
measurements on single crystals of the iron-based superconductor 
K$_{y}$Fe$_{2-x}$Se$_2$. The Knight shift demonstrates bulk singlet 
superconductivity. However, no coherence peak is observed under the measurement field. We find
Fermi-liquid behavior above $T_c$, which further indicates an absence 
of strong low-energy spin fluctuations on the Se site in the normal state. 
Our experimental evidence for singlet superconductivity with no coherence 
peak and no clear evidence of spin fluctuations on the Se site constitutes 
essential ingredients in formulating a complete understanding of this newest 
family of iron selenide superconductors. Such an understanding would in turn 
contribute to unravelling the mystery of high-$T_c$ superconductivity in 
general.   
  
We thank T. Li, Z. Y. Lu, B. Normand, S. C. Wang, X. Q. Wang, Y. Su, T. Xiang, 
G. M. Zhang, and Q. M. Zhang for stimulating discussions. Work at Renmin 
University of China was supported by the Natural Science Foundation of 
China (Grant No. 10974254, No. 11034012, and No. 11074304) and the National 
Basic Research Program of China (973 program) under contract 
No. 2010CB923004 and No. 2011CBA00112.   


\end{document}